\documentclass[11pt,reqno]{amsart}
\usepackage{amsfonts}

\newtheorem{thm}{Theorem}[section]
\newtheorem{cor}{Corollary}[section]

\newcommand{\mysection}[1]{\section{#1}\setcounter{equation}{0}}

\newfont{\bb}{msbm10 at 12pt}

\def\pf{{\textit {Proof :} }}

\def\R{\hbox{\bb R}}

\def\N{\mathcal N}

\newcommand{\bal}{\begin{align}}      \newcommand{\eal}{\end{align}}
\newcommand{\ba}{\begin{array}}      \newcommand{\ea}{\end{array}}
\newcommand{\bc}{\begin{center}}     \newcommand{\ec}{\end{center}}
\newcommand{\be}{\begin{enumerate}}  \newcommand{\ee}{\end{enumerate}}
\newcommand{\beq}{\begin{eqnarray}}  \newcommand{\eeq}{\end{eqnarray}}
\newcommand{\beQ}{\begin{eqnarray*}} \newcommand{\eeQ}{\end{eqnarray*}}
\newcommand{\bi}{\begin{itemize}}    \newcommand{\ei}{\end{itemize}}
\newcommand{\bt}{\begin{tabular}}    \newcommand{\et}{\end{tabular}}
\newcommand{\bdm}{\begin{displaymath}} \newcommand{\edm}{\end{displaymath}}

\newcommand{\rw}{\rightarrow}

\def\qed{\hfill{Q.E.D.}\smallskip}

\newcommand{\ls}{\setlength{\baselineskip}{12pt}
                 \setlength{\parskip}{3mm}}

\begin{document}

\title[ADM and Bondi energy-momenta II]{On the relation between ADM
and Bondi energy-momenta -- radiative spatial infinity}

\author{Wen-ling Huang}
\address[Wen-ling Huang]{Department Mathematik, Schwerpunkt GD, Universit{\"a}t Hamburg, Bundesstr.
55, D-20146 Hamburg, Germany} \email{huang@math.uni-hamburg.de}
\author{Xiao Zhang}
\address[Xiao Zhang]{Institute of Mathematics, Academy of Mathematics and
Systems Science, Chinese Academy of Sciences, Beijing 100080,
China}
\email{xzhang@amss.ac.cn}

\begin{abstract}
In a vacuum spacetime equipped with the Bondi's radiating metric
which is asymptotically flat at spatial infinity including
gravitational radiation ({\bf Condition D}), we establish the
relation between the ADM total linear momentum and the Bondi
momentum. The relation between the ADM total energy and the Bondi
mass in this case was established earlier in \cite{Z4}.
\end{abstract}



\maketitle \pagenumbering{arabic}

\mysection{Introduction}
 \ls

It is a fundamental problem in gravitational radiation what the
relation is between the ADM energy-momentum and the Bondi
energy-momentum. Ashtekar and Magnon-Ashtekar demonstrated the
mass at spatial infinity is the past limit of the Bondi mass taken
as the cut approaches the ``point'' of spatial infinity in the
frame of Penrose conformal compactification \cite{AM} (See also
the related works \cite{H, V1, V2}). It was proved the standard
ADM mass at spatial infinity is the past limit of the Bondi mass
for some strongly asymptotically flat, globally hyperbolic vacuum
\cite{CK}, and for vacuum spacetimes equipped with the Bondi's
radiating metric with ``strongly" asymptotic flatness at spatial
infinity ({\bf Condition C})\cite{Z4}. However, it is presumably
believed the assumed asymptotic flatness at spatial infinity in
all above works precludes gravitational radiation, at least near
spatial infinity.

Under some weaker assumption of asymptotic flatness at spatial
infinity ({\bf Condition D}), the second author also derive a
formula relating the ADM total energy to the Bondi mass. In this
case, the ADM total energy is no longer the past limit of the Bondi
mass and they differ by certain quantity relating to the {\it news}
of the system. The most importance is that gravitational radiation
should be included under this condition. Some physical
interpretation of {\bf Condition D} is provided in \cite{Z4}.

The main difference between {\bf Condition C} and {\bf Condition D}
is as follows: Under {\bf Condition C}, the initial data set
$\big\{t= t _0\big\}$ is standard asymptotically flat in the sense
that, in natural coordinates $\{x^u\}$, the metric $g$ and the
second fundamental form $h$ satisfy
 $$
g _{uv} =\delta _{uv}+O\Big(\frac{1}{r}\Big),
\partial _k g _{uv} = O\Big(\frac{1}{r^2}\Big),
\partial _l \partial _k g _{uv} = O\Big(\frac{1}{r^3}\Big),
 $$
 $$
h _{uv} = O\Big(\frac{1}{r^2}\Big),\partial _k h _{uv} =
O\Big(\frac{1}{r^3}\Big).
 $$
However, under {\bf Condition D}, the initial data set $\big\{t= t
_0\big\}$ is weaker asymptotically flat in the sense that, in
natural coordinates $\{x^u\}$, the metric $g$ and the second
fundamental form $h$ satisfy
 $$
g _{uv} = \delta _{uv} + O\Big(\frac{1}{r}\Big),
\partial _k g _{uv} = O\Big(\frac{1}{r}\Big),
\partial _l \partial _k g _{uv} = O\Big(\frac{1}{r}\Big),
 $$
 $$
h _{uv} = O\Big(\frac{1}{r}\Big), \partial _k h _{uv} =
O\Big(\frac{1}{r}\Big).
 $$

Note that an observer detecting gravitational waves locates always
in some finite region of a spacelike hypersurface. Thus the data the
observer detects is from the gravitational waves arrived at the
spacelike hypersurfaces. Suppose the metric and the second
fundamental form of this spacelike hypersurface can be read from the
data, then we can calculate the integrands of the ADM total
energy-momentum and integrate them in sphere with radius $r$
included in this region. One may wonder whether gravitational waves
can reach spatial infinity. However, as the ADM total
energy-momentum is defined as the limit $r \rw \infty$, the data
from gravitational waves should reach spatial infinity and {\bf
Condition D} should be reasonable when we calculate the ADM total
energy-momentum.

In this paper, we will study the relation between the ADM total
energy-momentum and the Bondi energy-momentum under {\bf Condition
D}. Although the second fundamental form $h$ of t-slice falls off as
slowly as $O(\frac{1}{r})$, $h- tr_g (h)g$ falls off surprisingly as
$O(\frac{1}{r ^2})$ at spatial infinity due to certain mysterious
cancelation. Therefore, the ADM total linear momentum is still
finite. It is unclear whether the ADM total energy-momentum is a
geometric invariant, i.e., independent on the choice of coordinates,
under {\bf Condition D}, however, we compare the ADM and Bondi
energy-momenta in a fixed coordinate. We believe it should have
physical meaning.
\mysection{Bondi's radiating vacuum spacetimes} \ls

The Bondi's radiating vacuum spacetime $\big(L ^{3,1}, \tilde
g\big) $ is a vacuum spacetime equipped with the following metric
$\tilde g =\tilde g _{ij} dx ^i dx ^j$
 \beq
\tilde g &=&\Big(\frac{V}{r} e ^{2\beta} +r ^2 e ^{2 \gamma} U ^2
\cosh 2\delta +r ^2 e ^{-2 \gamma} W ^2 \cosh 2\delta \nonumber\\
&&+2 r ^2 UW \sinh 2 \delta \Big)du ^2 -2e ^{2\beta}
du dr   \nonumber\\
& &-2r ^2 \Big(e ^{2 \gamma} U \cosh 2\delta +W \sinh 2 \delta
\Big) du d\theta     \nonumber\\
& &-2r ^2 \Big(e ^{-2 \gamma} W \cosh 2\delta+U \sinh 2\delta
\Big)\sin \theta du d\phi    \nonumber\\
& &+r ^2 \Big(e ^{2 \gamma} \cosh 2\delta d\theta ^2 +e ^{-2
\gamma}\cosh 2\delta \sin ^2 \theta d \psi ^2   \nonumber\\
& &+2 \sinh 2\delta \sin \theta d \theta d \psi \Big),
\label{bondi-metric}
 \eeq
where $\beta, \gamma, \delta, U, V, W$ are functions of
 \beQ
x ^0 =u,\;\;x ^1=r,\;\;x ^2=\theta, \;\;x ^3=\psi,
 \eeQ
$u$ is a retarded coordinate, $r$ is Euclidean distance, $\theta $
and $\psi $ are spherical coordinates, $0 \leq \theta \leq \pi$,
$0 \leq \psi \leq 2 \pi$. We assume that $\tilde g$ satisfies {\it
the outgoing radiation condition}.

The metric (\ref{bondi-metric}) was studied by Bondi, van der
Burg, Metzner and Sachs in the theory of gravitational waves in
general relativity \cite{BBM, S, vdB}. They proved that the
following asymptotic behavior holds for $r$ sufficiently large if
the spacetime satisfies the outgoing radiation condition
\cite{vdB}
 \beQ
\gamma &=&\frac{c(u, \theta, \psi)}{r} +\frac{C(u,\theta,
\psi)-\frac{1}{6} c ^3 -\frac{3}{2} c d ^2}{r ^3}
+O\Big(\frac{1}{r ^4}\Big),  \\
\delta &=&\frac{d(u, \theta, \psi)}{r} +\frac{H(u,\theta,
\psi)+\frac{1}{2} c ^2 d -\frac{1}{6} d ^3}{r ^3}
+O\Big(\frac{1}{r ^4}\Big),  \\
\beta &=&-\frac{c ^2 + d ^2}{4r ^2} +O\Big(\frac{1}{r ^4}\Big),     \\
U &=& -\frac{l(u, \theta, \psi)}{r ^2} +\frac{p(u, \theta,
\psi)}{r ^3}
+O\Big(\frac{1}{r ^4}\Big),  \\
W &=& -\frac{\bar l(u, \theta, \psi)}{r ^2}+\frac{\bar p(u,
\theta, \psi)}{r ^3}+O\Big(\frac{1}{r ^4}\Big),  \\
V &=& -r +2 M (u, \theta, \psi)+\frac{\bar M(u, \theta, \psi)}{r}
+O\Big(\frac{1}{r^2}\Big),
 \eeQ
where
 \beQ
l &=& c _{, 2} +2c \cot \theta +d _{, 3} \csc \theta,\\
\bar l &=& d _{, 2} +2d \cot \theta -c _{,3} \csc \theta,\\
p &=& 2N +3(c c _{, 2}+d d _{, 2}) +4(c ^2 +d ^2)\cot \theta\\
  & &-2(c_{,3} d -c d _{,3}) \csc \theta,\\
\bar p &=& 2P +2(c _{, 2} d -c d _{, 2}) +3(c c_{,3} +d d
_{,3})\csc \theta,\\
\bar M &=&N _{,2} +\cot \theta +P _{, 3} \csc \theta
-\frac{c ^2 +d ^2}{2}\\
& &-\big[(c _{,2}) ^2 +(d _{,2}) ^2 \big]-4(c c _{,2} +d d
_{,2})\cot \theta \\
& & -4(c ^2 +d ^2) \cot ^2
\theta -\big[(c _{,3}) ^2 +(d _{,3}) ^2 \big]\csc ^2 \theta \\
& &+4(c _{,3} d -c d _{, 3}) \csc \theta \cot \theta +2(c_{,3} d
_{,2}-c _{,2} d _{,3})\csc \theta.
 \eeQ
(We denote $f _{,i} = \frac{\partial f}{\partial x ^i }$ for
$i=0,1,2,3$ throughout the paper.) $M$ is the {\it mass aspect}
and $c_{,0}$, $d _{, 0}$ are the {\it news functions} and they
satisfy the following equation \cite{vdB}:
 \beq
M _{,0} =-\Big[(c _{,0} )^2 +(d _{, 0} )^2 \Big]
+\frac{1}{2}\Big(l _{,2} +l \cot \theta +\bar l _{,3} \csc \theta
\Big) _{,0} .\label{u-deriv}
 \eeq

Let $N _{u _0}$ be a null hypersurface which is given by $u=u _0$
at null infinity. The Bondi energy-momentum of $N _{u _0}$ is
defined by \cite{BBM}:
 \beQ
m _\nu (u _0) = \frac{1}{4 \pi} \int _{S ^2} M (u _0, \theta,
\psi) n ^{\nu} d S
 \eeQ
where $\nu =0, 1, 2, 3$, $S ^2 $ is the unit sphere,
 \beQ
n ^0 =1,\;\; n ^1 = \sin \theta \cos \psi,\;\; n ^2 = \sin \theta
\sin \psi,\;\; n ^3 = \cos \theta.
 \eeQ
And $m _0$ is the Bondi mass, $m _i$ is the Bondi momentum.

Denote by $\{ \breve{e} ^i \}$ the coframe of the standard flat
metric $g _0$ on $\R ^3$ in polar coordinates,
 \beQ
\breve{e} ^1 = dr, \;\;\breve{e} ^2 = r d\theta, \;\;\breve{e} ^3
= r \sin \theta d\psi.
 \eeQ
Denote by $\{\breve{e} _i \}$ the dual frame ($i=1,2,3$). The
connection 1-form $\{\breve{\omega } _{ij}\}$ is defined by
 \beQ
d \breve{e} ^i= - \breve{\omega} _{ij} \wedge \breve{e} ^{j}.
 \eeQ
It is easy to find that
 \beQ
    \breve{\omega}  _{12} =  -\frac{1}{r}\breve{e}  ^{2}, \;\;
    \breve{\omega}  _{13} =  -\frac{1}{r}\breve{e}  ^{3}, \;\;
    \breve{\omega}  _{23} =  -\frac{\cot \theta}{r}\breve{e} ^{3}.
 \eeQ
The Levi-Civita connection $\breve{\nabla}$ of $g _0$ is given by
 \beQ
\breve{\nabla} \breve{e} _{i}= - \breve{\omega} _{ij} \otimes
\breve{e} _{j}.
 \eeQ
We denote $\breve{\nabla} _i \equiv \breve{\nabla} _{\breve{e}
_i}$ for $i=1,2,3$ throughout the paper.

Define $\mathcal{C} _{\{a_1, a_2, a_3\}}$ the space of smooth
functions in the spacetime which satisfies the following
asymptotic behavior at spatial infinity
 \beq
\mathcal{C} _{\{a_1, a_2, a_3\}} =
 \left\{f:
 \begin{array}{ccc}
    \lim _{r \rightarrow \infty}\lim _{u \rightarrow -\infty} r ^{a_1}f
    &=&O\big(1\big),\\
    \lim _{r \rightarrow \infty}\lim _{u \rightarrow -\infty} r ^{a_2}
    \breve{\nabla} _i f
    &=&O\big(1\big),\\
    \lim _{r \rightarrow \infty}\lim _{u \rightarrow -\infty} r ^{a_3}
    \breve{\nabla} _i\breve{\nabla} _j f &=& O\big(1\big)
 \end{array}
 \right\}.       \label{mathcal-C}
 \eeq
In \cite{Z4}, the following four conditions are introduced:
 \begin{description}
 \item[Condition A] {\bf Each of the six functions $\beta $,
$\gamma$, $\delta$, $U$, $V$, $W$ together with its derivatives up
to the second orders are equal at $\psi =0$ and $2\pi$}.\\
 \item[Condition B] {\bf For all $u$,
 \beQ
\int _0 ^{2\pi} c(u, 0, \psi) d\psi =0, \;\; \int _0 ^{2\pi} c(u,
\pi, \psi) d\psi =0.
 \eeQ}
 \item[Condition C] $\;\;\gamma \in \mathcal{C} _{\{1, 2, 3\}},\;\;
\delta \in \mathcal{C} _{\{1, 2, 3\}}, \;\;\beta \in \mathcal{C}
_{\{2, 3, 4\}}, \;\;U \in \mathcal{C} _{\{2, 3, 4\}}, \;\;W \in
\mathcal{C} _{\{2, 3, 4\}}, \;\;V+r \in \mathcal{C} _{\{0, 1,
2\}}$.\\
 \item[Condition D] $\;\;\gamma \in \mathcal{C} _{\{1, 1,
1\}},\;\; \delta \in \mathcal{C} _{\{1, 1, 1\}}, \;\;\beta \in
\mathcal{C} _{\{2, 2, 2\}}, \;\;U \in \mathcal{C} _{\{2, 2, 2\}},
\;\;W \in \mathcal{C} _{\{2, 2, 2\}}, \;\;V+r \in \mathcal{C}
_{\{0, 0, 0\}}$.\\
 \end{description}

{\bf Condition A} and {\bf Condition B} ensure that the metric
(\ref{bondi-metric}) is regular, also ensure the following Bondi
mass loss formula
 \beQ
\frac{d}{du} m _\nu =-\frac{1}{4 \pi} \int _{S ^2} \big[(c _{,0})
^2 +(d _{,0}) ^2 \big] n ^\nu d S.
 \eeQ
{\bf Condition C} ensures the Schoen-Yau's positive mass theorem at
spatial infinity. However, it precludes gravitational radiation, at
least near spatial infinity. {\bf Condition D} should include
gravitational radiation. It indicates that, for $r$ sufficiently
large,
 \beQ
\lim _{u \rightarrow -\infty} \Big\{M, c, d, M _{,0}, c _{,0}, d
_{,0}, M _{,A}, c _{,A}, d _{,A} \Big\}=O\big(1\big)
 \eeQ
where $2 \leq A\leq 3$.
\mysection{Initial data sets} \ls

From now on we assume the ``real'' time $t$ is defined as
 \beQ
t=u+r.
 \eeQ
An initial data set $\big(N _{t _0}, g, h\big)$ is a spacelike
hypersurface in $L ^{3,1}$ which is given by $\big\{t= t
_0\big\}$. Here $g$ is the induced metric of $\tilde g$ and $h$ is
the second fundamental form. It is straightforward that
 \beq
g&=&\Big(\big(2+\frac{V}{r}\big) e ^{2\beta} +r ^2 e ^{2 \gamma} U
^2 \cosh 2\delta    \nonumber\\
& &+r ^2 e ^{-2 \gamma} W ^2 \cosh 2\delta +2 r ^2 UW \sinh 2
\delta  \Big) dr ^2   \nonumber\\
& & +r ^2 \Big(e ^{2 \gamma} \cosh 2\delta d\theta ^2
 +e ^{-2\gamma}\cosh 2\delta \sin ^2 \theta d \psi ^2 \nonumber\\
 & & +2 \sinh 2\delta \sin \theta d \theta d \psi \Big)
 \nonumber\\
& &+2 r ^2 \Big(e ^{2 \gamma } U \cosh 2 \delta +W \sinh 2 \delta
\Big)dr d\theta     \nonumber\\
& &+2 r ^2 \Big(e ^{-2 \gamma } W \cosh 2 \delta +U \sinh 2 \delta
\Big)\sin \theta dr d\psi . \label{metric}
 \eeq
The other components of the metric $\tilde g$ are
 \beQ
\tilde g _{tt} &=& \Big(\frac{V}{r} e ^{2\beta} +r ^2 e ^{2 \gamma}
U ^2 \cosh 2\delta   \\ & &+r ^2 e ^{-2 \gamma} W ^2 \cosh
2\delta +2 r ^2 UW \sinh 2 \delta\Big)\\
 \tilde g_{t1}
&=&-\Big(1+\frac{V}{r}\Big) e ^{2 \beta} -r ^2 e ^{2 \gamma} U ^2
\cosh 2\delta \\ & &-r ^2 e ^{-2 \gamma} W ^2 \cosh 2\delta -2 r
^2 UW \sinh 2 \delta, \\
 \tilde g_{t2}&= & - r ^2 \Big(e ^{2 \gamma } U
\cosh 2 \delta +W \sinh 2 \delta \Big),\\
 \tilde g_{t3}&= & - r ^2 \Big(e
^{-2 \gamma } W \cosh 2 \delta +U \sinh 2 \delta \Big)\sin \theta.
 \eeQ
The inverse $g ^{ij}$ of metric tensor $g _{ij}$, the lapse $\N
=\Big(-\tilde g ^{tt} \Big) ^{-\frac{1}{2}}$ and the shift $X _i
=\tilde g _{ti}$ $(i=1,2,3)$ of $N _{t_0}$ are derived in
\cite{Z4} as follows:
 \beQ
\frac{1}{g ^{11}}&=&\big(2+\frac{V}{r}\big) e ^{2\beta},\\
g ^{12} &=& -U g ^{11},\\
g ^{13} &=& -\frac{W}{\sin \theta} g ^{11},\\
g ^{22} &=& \frac{e ^{-2 \gamma} \cosh 2\delta}{r ^2} +U ^2 g^{11},\\
g ^{23} &=& -\frac{ \sinh 2\delta}{r ^2 \sin \theta}
+\frac{UW}{\sin \theta} g^{11},\\
 g ^{33} &=& \frac{e ^{2 \gamma}
\cosh 2\delta}{r ^2 \sin ^2 \theta} +\frac{W ^2}{\sin ^2 \theta}
g^{11},\\
 \N ^2&=&e ^{4 \beta} g ^{11},\\
X _1 &=&-\Big(1+\frac{V}{r}\Big) e ^{2 \beta} -r ^2 e ^{2 \gamma} U
^2 \cosh 2\delta\\
 & &-r ^2 e ^{-2 \gamma} W ^2 \cosh 2\delta -2 r
^2 UW \sinh 2
\delta,\\
 X _2&= & - r ^2 \Big(e ^{2 \gamma } U \cosh 2 \delta +W
\sinh 2 \delta
\Big),\\
X _3&= & - r ^2 \Big(e ^{-2 \gamma } W \cosh 2 \delta +U \sinh 2
\delta \Big)\sin \theta.
 \eeQ
The second fundamental form is then given by
 \beQ
h _{ij} =\frac{1}{2 \N} \Big(\nabla _i X _j +\nabla _j X _i
-\partial _t \tilde g _{ij} \Big) _{t=t_0}.
 \eeQ

With the help of asymptotic behavior of $\beta, \gamma, \delta, U,
V, W$ and Mathematica 5.0, we obtain the asymptotic expansions of $g
_{ij}$,
 \beQ
g _{11}& =&1 + \frac{2M}{r}+\frac{1}{r^2}\Big(-\frac{{c}^2}{2}-
\frac{{d}^2}{2} + l ^2 +\bar l ^2 +\bar M \Big)+O\Big(\frac{1}{r ^3}\Big),\\
 g _{22}& =&r^2 + 2r c + 2 {c}^2 + 2 {d}^2+O\Big(\frac{1}{r}\Big),\\
g _{33}& =&\Big(r^2  - 2 r c  + 2 {c}^2  +
  2 {d}^2 \Big){\sin ^2 \theta}+O\Big(\frac{1}{r}\Big),\\
g _{12}&=&-l + \frac{1}{r}\Big(-2c l -2 d \bar l +p\Big)+O\Big(\frac{1}{r ^2}\Big),\\
g _{13}&=&-\bar l \sin \theta + \frac{\sin \theta}{r}\Big(2c \bar l -2 d l +\bar p\Big)+O\Big(\frac{1}{r ^2}\Big),\\
g _{23}&=& 2r d \sin \theta +O\Big(\frac{1}{r}\Big),
 \eeQ
the asymptotic expansions of $g ^{ij}$,
 \beQ
 g ^{11}&=& 1 -
\frac{2M}{r} + \frac{1}{r^2}\Big(\frac{{c}^2}{2}+
\frac{{d}^2}{2} + 4M ^2 -\bar M \Big)+O\Big(\frac{1}{r^3}\Big),\\
 g ^{22} &=&\frac{1}{r ^2} - \frac{2c}{r^3} +
\frac{1}{r ^4} \Big(2{c}^2 + 2{d}^2 +l ^2 \Big)+O\Big(\frac{1}{r^5}\Big),\\
g ^{33} &=&\frac{\csc ^2 \theta }{r^2} + \frac{2c \csc ^2 \theta}{r
^3} +\frac{\csc ^2 \theta}{r ^4}\Big(2{c}^2 + 2{d}^2 +\bar l ^2\Big)+O\Big(\frac{1}{r^5}\Big),\\
 g ^{12} &=&\frac{l}{r ^2}-\frac{1}{r ^3}\Big(2M l +p\Big)+O\Big(\frac{1}{r^4}\Big),\\
g ^{13} &=&\frac{\bar l \csc \theta }{r ^2}-
  \frac{\csc \theta}{r ^3}\Big(2M \bar l +\bar p\Big)+O\Big(\frac{1}{r ^4}\Big)\\
g ^{23} &=&-\frac{2d \csc \theta}{r^3}
   +\frac{l \bar l \csc \theta}{r ^4}+O\Big(\frac{1}{r^5}\Big),
 \eeQ
and the asymptotic expansions of $N$ and $X_i$,
 \beQ
\N &=&1 - \frac{M}{r} - \frac{1}{r^2}\Big(\frac{{c}^2}{4} +
 \frac{{d}^2}{4}-\frac{3 M ^2}{2}+\frac{\bar M}{2}\Big)+O\Big(\frac{1}{r^3}\Big), \\
X _1 &=&-\frac{2 M}{r} -\frac{1}{r ^2}\Big(l ^2 +\bar l ^2 +\bar M
\Big)+O\Big(\frac{1}{r^3}\Big),\\
X _2 &=&l +
  \frac{1}{r}\Big(2cl +2d \bar l -p\Big)+O\Big(\frac{1}{r^2}\Big),\\
X _3 &=&\bar l \sin \theta + \frac{\sin \theta}{r}\Big(-2c \bar l
+2d l -\bar p \Big)+O\Big(\frac{1}{r^2}\Big).
 \eeQ
as well as the asymptotic expansions of $h _{ij}$,
 \beQ
 h _{11}& =&\frac{M _{,0}}{r}+\frac{1}{r ^2}\Big(2M -M M _{,0}
 +\frac{c c_{,0}}{2}+\frac{d d_{,0}}{2} +l l _{,0} +\frac{\bar M _{,0}}{2}\Big)+O\Big(\frac{1}{r ^3}\Big),\\
 h _{22}& =& -r c _{,0}-2 M+M c _{,0} -2c c _{,0} -2 d d_{,0}+l
 _{,2}+O\Big(\frac{1}{r}\Big),\\
 h _{33}& =&\Big(r c_{,0} -2 M  - M c _{,0} -2 c c _{,0}
 -2 d d_{,0} +l \cot \theta + \bar l
 _{,3} \csc \theta\Big)\sin ^2 \theta \\
 & &+O\Big(\frac{1}{r}\Big),\\
 h _{12}&=&\frac{1}{r}\Big(- M_{,2} -l + c_{,0} l+d _{,0} \bar l\Big)+O\Big(\frac{1}{r ^2}\Big),\\
 h _{13}&=&\frac{\sin \theta}{r}\Big(-M_{,3}\csc \theta -\bar l  - c_{,0} \bar l   + d _{,0}l \Big)
+O\Big(\frac{1}{r ^2}\Big) ,\\
 h _{23}&=&\Big[-
  r d_{,0}+M d_{,0}+\frac{1}{2}\big( \bar l _{,2}-\bar l
  \cot \theta +l _{,3} \csc \theta \big)\Big]\sin \theta+O\Big(\frac{1}{r}\Big) .
 \eeQ
The trace of the second fundamental form is
 \beQ
tr _g \big(h\big)&=&\frac{M_{,0}}{r} + \frac{1}{r^2}\Big( -2M -3M M
_{,0} +l _{,2} +l \cot \theta +\bar
l _{,3} \csc\theta \\
& &+\frac{c c_{,0}}{2}+\frac{d d_{,0}}{2} +l l _{,0} +\frac{\bar M
_{,0}}{2} \Big) +O\Big(\frac{1}{r ^3}\Big).
 \eeQ
\mysection{ADM and Bondi total momenta} \ls

Let Euclidean coordinates
 \beQ
y ^1 =r \sin \theta \cos \psi,\;\;\;\; y ^2=r \sin \theta \sin
\psi, \;\;\;\;y ^3 =r \cos \theta.
 \eeQ
In polar coordinates, the ADM total energy $\mathbb{E}$ and the
ADM total linear momentum $\mathbb{P} _k$ are \cite{ADM, Z4}
 \beQ
 \mathbb{E} &=&\frac{1}{16\pi}\lim _{r \rightarrow \infty} \int
_{S _r} \Big[\breve {\nabla } ^j g\big(\breve{e} _1, \breve{e}
_j\big) -\breve {\nabla } _1 tr _{g _0} \big(g\big) \Big]
\breve{e} ^2 \wedge \breve{e} ^3,\\
 \mathbb{P} _k &=&\frac{1}{8\pi}\lim _{r \rightarrow
\infty} \int _{S _r} \Big[h\Big(\frac{\partial}{\partial y ^k},
\frac{\partial}{\partial r}\Big) -g\Big(\frac{\partial}{\partial y
^k}, \frac{\partial}{\partial r}\Big)tr _{g} \big(h\big) \Big]
\breve{e} ^2 \wedge \breve{e} ^3.
 \eeQ
Under {\bf Condition A}, {\bf Condition B} and {\bf Condition D},
the second author derived the relation between the ADM total
energy and the Bondi mass \cite{Z4},
 \beQ
\mathbb{E} (t _0) = m _0 (-\infty) +\frac{1}{2\pi}\lim
_{u\rightarrow -\infty} \int _0 ^{\pi} \int _0 ^{2\pi} \Big(c c
_{,0} +d d _{,0} \Big) \sin\theta d\psi d\theta.
 \eeQ
Now we study the relation between the ADM total linear momentum
and the Bondi momentum under these conditions and prove the main
theorem.
 \begin{thm}\label{momentum}
Let $\mathbb{P} _k (t _0)$ be the ADM total linear momentum of
spacelike hypersurface $N _{t_0}$ whose metric satisfies
(\ref{metric}). Under {\bf Condition A}, {\bf Condition B} and
{\bf Condition D}, we have
 \beQ
\mathbb{P} _k (t_0)= m _k (-\infty)+\frac{1}{8\pi}\lim _{u
\rightarrow -\infty} \int _0 ^{\pi} \int _0 ^{2 \pi} {\mathcal P}
_k d \psi d \theta
 \eeQ
for $k=1,2,3$, where
 \beQ
{\mathcal P} _1 &=&\Big[\big(c _{,0}+M _{,0}\big) \cos \theta
\cos \psi -d_{,0} \sin \psi\Big]l \sin \theta\\
             & &+\Big[\big(c _{,0}-M _{,0}\big)
\sin \psi +d_{,0} \cos \theta \cos \psi\Big]\bar l \sin \theta,\\
{\mathcal P} _2 &=&\Big[\big(c _{,0}+M _{,0}\big) \cos \theta
\sin \psi +d_{,0} \cos \psi\Big]l \sin \theta \\
             & &-\Big[\big(c _{,0}-M _{,0}\big)
\cos \psi -d_{,0} \cos \theta \sin \psi\Big]\bar l \sin \theta,\\
{\mathcal P} _3 &=&-\big(c _{,0}+M _{,0}\big)l \sin ^2 \theta  -
             d _{,0}\bar l \sin ^2 \theta.
 \eeQ
 \end{thm}
\pf Using the asymptotic expansions of $g _{ij}$ and $h _{ij}$, we
obtain
 \beQ
g\Big(\frac{\partial}{\partial y ^1}, \frac{\partial}{\partial r}
\Big)&=&g _{11} n ^1 +g _{21} \frac{\cos \theta \cos \psi}{r}-g
_{31}
\frac{\sin \psi}{r \sin \theta}\\
&=&\sin \theta \cos \psi+ \frac{1}{r}\Big(2M \sin \theta\cos \psi\\
& &-l \cos\theta \cos \psi +\bar l \sin \psi \Big)+O\Big(\frac{1}{r ^2}\Big),\\
g\Big(\frac{\partial}{\partial y ^2}, \frac{\partial}{\partial r}
\Big)&=&g _{11} n ^2 +g _{21} \frac{\cos \theta \sin \psi}{r}+g
_{31} \frac{\cos \psi}{r \sin \theta}\\
&=&\sin \theta \sin \psi + \frac{1}{r}\Big(2 M \sin \theta
\sin \psi \\
& &-l \cos \theta \sin \psi -\bar l \cos \psi \Big)+O\Big(\frac{1}{r ^2}\Big) ,\\
g\Big(\frac{\partial}{\partial y ^3}, \frac{\partial}{\partial r}
\Big)&=&g _{11} n ^3 -g _{21} \frac{\sin \theta}{r}\\
&=&\cos \theta +\frac{1}{r} \Big(2M \cos \theta +l \sin \theta
\Big)+O\Big(\frac{1}{r ^2}\Big),\\
 h\Big(\frac{\partial}{\partial y ^1}, \frac{\partial}{\partial r}
\Big)&=&h _{11} n ^1 +h _{21} \frac{\cos \theta \cos \psi}{r}-h
_{31} \frac{\sin \psi}{r \sin \theta}\\
&=&\frac{M_{,0}\sin \theta \cos \psi}{r} +
  \frac{1}{r ^2} \Big[\big(2M -M M _{,0}\\
  & &+\frac{c c_{,0}}{2}+\frac{d d_{,0}}{2} +l l _{,0} +\frac{\bar M
  _{,0}}{2}\big)\sin \theta \cos \psi \\
  & &-\big(M _{,2} +l -c _{,0}l -d
  _{,0} \bar l\big)\cos \theta \cos \psi \\
  & &+\big(M _{,3}\csc \theta
  +\bar l +c _{,0} \bar l -d _{,0} l\big)\sin \psi \Big]+O\Big(\frac{1}{r
     ^3}\Big),\\
 h\Big(\frac{\partial}{\partial y ^2}, \frac{\partial}{\partial r}
\Big)&=&h _{11} n ^2 +h _{21} \frac{\cos \theta \sin \psi}{r}+h
_{31} \frac{\cos \psi}{r \sin \theta}\\
&=&\frac{M_{,0}\sin \theta \sin \psi}{r} +
  \frac{1}{r ^2}\Big[\big(2M -M M _{,0}\\
  & &+\frac{c c_{,0}}{2}+\frac{d d_{,0}}{2}
  +l l _{,0} +\frac{\bar M
  _{,0}}{2}\big)\sin \theta \sin \psi \\
  & &-\big(M _{,2} +l -c _{,0}l -d
  _{,0} \bar l\big)\cos \theta \sin \psi \\
  & &-\big(M _{,3}\csc \theta
  +\bar l +c _{,0} \bar l -d _{,0} l\big)\cos \psi \Big]+O\Big(\frac{1}{r
     ^3}\Big),\\
 h\Big(\frac{\partial}{\partial y ^3}, \frac{\partial}{\partial r}
\Big)&=&h _{11} n ^3 -h _{21} \frac{\sin \theta}{r}\\
&=&\frac{M _{,0}\cos \theta}{r} +
  \frac{1}{r ^2}\Big[\big(2M -M M _{,0}\\
  & &+\frac{c c_{,0}}{2}+\frac{d d_{,0}}{2} +l l _{,0} +\frac{\bar M
  _{,0}}{2}\big)\cos \theta \\
  & & +\big(M _{,2} +l -c _{,0}l -d
  _{,0} \bar l\big)\sin \theta \Big]+O\Big(\frac{1}{r
     ^3}\Big).
 \eeQ
Denote $\mathbb{K} _k =h\Big(\frac{\partial}{\partial y ^k},
\frac{\partial}{\partial r}\Big) -g\Big(\frac{\partial}{\partial y
^k}, \frac{\partial}{\partial r}\Big)tr _{g} \big(h\big)$. We then
obtain
 \beQ
\mathbb{K} _1&=&\frac{1}{r ^2}\Big\{4M \sin \theta \cos \psi -M
_{,2}
\cos \theta \cos \psi +M _{,3} \csc \theta \sin \psi \\
& &- l _{,2} \sin \theta \cos \psi-\bar l _{,3} \cos \psi \\
& &+ \big[( c _{,0}+M _{,0}-2) \cos \theta \cos \psi -d _{,0} \sin
\psi  \big] l \\
& & + \big[(c _{,0} -M _{,0}+1) \sin \psi +d _{,0} \cos \theta \cos
\psi \big]\bar l \Big\}+O\Big(\frac{1}{r ^3}\Big),\\
\mathbb{K} _2&=&\frac{1}{r ^2}\Big\{4M \sin \theta \sin \psi -M
_{,2}
\cos \theta \sin \psi -M _{,3} \csc \theta \cos \psi\\
& &- l _{,2} \sin \theta \sin \psi -\bar l _{,3} \sin
\psi\\
& &+\big[( c _{,0} +M _{,0}-2)\cos \theta \sin \psi +d _{,0} \cos
\psi  \big]l \\
& & - \big[(c _{,0} -M _{,0}+1) \cos \psi -d _{,0} \cos \theta \sin
\psi \big]\bar l \Big\}+O\Big(\frac{1}{r ^3}\Big),\\
 \mathbb{K} _3&=&\frac{1}{r ^2}\Big\{4M \cos \theta  -M _{,2}
\sin \theta - l _{,2} \cos \theta -\bar l _{,3} \cot \theta  \\
& &-\big[( c _{,0} +M _{,0}-2)\sin \theta  +\csc \theta \big]l -d
_{,0} \bar l \sin \theta\Big\}+O\Big(\frac{1}{r ^3}\Big).
 \eeQ
Integrating $\mathbb{K} _k$ over $S _r$, and simplifying them by the
integration by part, also noting that for fixed $t=t _0$, $r
\rightarrow \infty$ is equivalent to $u \rightarrow -\infty$, we
finally obtain the proof of the theorem.\qed
\mysection{Axi-Symmetric spacetimes} \ls

Gravitational waves in axi-symmetric spacetimes were first studied
by Bondi, van der Burg and Metzner \cite{BBM}. In this case the
metric of the spacetime is
 \beq
\tilde g &=&\Big(\frac{V}{r} e ^{2\beta} +r ^2 e ^{2 \gamma} U ^2
\Big) du ^2 -2e ^{2\beta} du dr \nonumber\\ & &-2r ^2 e ^{2 \gamma}
U du d\theta +r ^2 \Big(e ^{2 \gamma} d\theta ^2 +e ^{-2 \gamma}
\sin ^2 \theta d \psi ^2 \Big) \label{bondi-metric-1}
 \eeq
where $\beta $, $\gamma $, $U$ and $V$ are functions of $u$, $r$
and $\theta$. This implies that
 \beQ
c=c(u,\theta), \;\;d=0,\;\; l =c _{,2} +2c \cot \theta,\;\;\bar l
=0.
 \eeQ
 \begin{cor}
Let $\mathbb{E} (t _0)$, $\mathbb{P} _k (t _0)$ be the ADM total
energy and the ADM total linear momentum of spacelike hypersurface
$N _{t_0}$ in axi-symmetric vacuum radiating spacetimes
(\ref{bondi-metric-1}). Under {\bf Condition A}, {\bf Condition B}
and {\bf Condition D}, we have
 \beQ
 \mathbb{E} (t _0) &=&m _0 (-\infty)+\lim _{u \rightarrow -\infty}
\int _0 ^{\pi} c c_{,0} \sin \theta d \theta,\\
 \mathbb{P} _1 (t _0)&=& m _1(-\infty),\\
 \mathbb{P} _2 (t _0)&=& m _2(-\infty),\\
 \mathbb{P} _3 (t_0) &=& m _3 (-\infty)-\frac{1}{4}\lim _{u \rightarrow -\infty}
\int _0 ^{\pi} \Big[-(c _{,0})^2 +\frac{c _{,220}}{2}\\
& & +c _{,20} \cot \theta \big)\Big] \big(c _{,2} +2c \cot \theta
\big) \sin ^2 \theta d \theta.
 \eeQ
 \end{cor}

\mysection{Appendix} \ls

In this appendix, we provide an addendum to \cite{Z4}. There are two
sign differences between the Bondi radiating metric $(2.1)$ in
\cite{Z4} and the metric used in current paper as well as the
original paper \cite{BBM}. They result some sign changes in certain
formulas but the main theorems in \cite{Z4} still holds true without
any change. We give as follows the corresponding formulas for
\cite{Z4} with respect to the metric used in current paper.

1) page 263, $(2.1)$ in \cite{Z4} changes to (\ref{bondi-metric}) in
current paper.

2) page 267, $(2.5)$ in \cite{Z4} changes to
 \beQ
\tilde g &=&\Big(\frac{V}{r} e ^{2\beta} +r ^2 e ^{2 \gamma} U ^2
\cosh 2\delta \\
 & &+r ^2 e ^{-2 \gamma} W ^2 \cosh 2\delta +2 r ^2 U W
\sinh 2 \delta\Big)dt ^2 \\
 & & -2\Big(\big(1+\frac{V}{r}\big) e
^{2\beta} +r ^2 e ^{2 \gamma} U ^2 \cosh 2\delta \\
 & &+r ^2 e ^{-2
\gamma} W ^2 \cosh 2\delta +2 r ^2 UW \sinh 2 \delta \Big) dt dr \\
 & &-2 r ^2 \Big(e ^{2 \gamma } U \cosh 2 \delta +W \sinh 2 \delta
\Big)dt d\theta \\
 & &-2 r ^2 \Big(e ^{-2 \gamma } W \cosh 2 \delta
+U
\sinh 2 \delta \Big)\sin \theta dt d\psi  \\
 & & +\Big(\big(2+\frac{V}{r}\big) e ^{2\beta} +r ^2 e ^{2 \gamma}
U ^2 \cosh 2\delta \\
& & +r ^2 e ^{-2 \gamma} W ^2 \cosh 2\delta +2
r
^2 UW \sinh 2 \delta \Big) dr ^2 \\
 & &+r ^2 \Big(e ^{2 \gamma} \cosh 2\delta d\theta ^2 +e
^{-2\gamma}\cosh 2\delta \sin ^2 \theta d \psi ^2 \\
 & &+2 \sinh2\delta \sin \theta d \theta d \psi \Big)\\
 & &+2 r ^2 \Big(e ^{2 \gamma } U \cosh 2 \delta +W \sinh 2 \delta
\Big)dr d\theta    \\
 & &+2 r ^2 \Big(e ^{-2 \gamma } W \cosh 2 \delta +U \sinh 2 \delta
\Big)\sin \theta dr d\psi.
 \eeQ

3) page 268, $(3.1)$ in \cite{Z4} change to (\ref{metric}) in
current paper.

4) page 269, line -9 to line -8 in \cite{Z4} changes to
 \beQ
g\big(\breve{e} _1, \breve{e} _1\big)&=&\Big[\big(2+\frac{V}{r}\big)
e
^{2\beta} +r ^2 e ^{2 \gamma} U ^2 \cosh 2\delta\\
& &+r ^2 e ^{-2 \gamma} W ^2 \cosh 2\delta +2 r ^2 UW \sinh 2 \delta
\Big] _{t=t _0} .
 \eeQ

5) page 271, line 2 to line 3 in \cite{Z4} change to
 \beQ
\frac{1}{g ^{11}}=\big(2+\frac{V}{r}\big) e ^{2\beta},
 \eeQ

6) page 271, line -7 to line -6 in \cite{Z4} change to
 \beQ
X _1 &=& -\Big(1+\frac{V}{r}\Big) e ^{2 \beta} -r ^2 e ^{2 \gamma}
U ^2 \cosh 2\delta \\
& &-r ^2 e ^{-2 \gamma} W ^2 \cosh 2\delta -2 r ^2 UW \sinh 2 \delta
 \eeQ

\hspace{2cm}

{\footnotesize {\it Acknowledgements.} {This work was partially done
when Wen-ling Huang visited the Morningside Center of Mathematics,
Chinese Academy of Sciences, and she would like to thank the center
for its hospitality. Research of Xiao Zhang is partially supported
by National Natural Science Foundation of China under grants
10231050, 10421001 and the Innovation Project of Chinese Academy of
Sciences.}}

\end{document}